\documentclass[10pt,conference]{IEEEtran}
\usepackage{cite}
\usepackage{amsmath,amssymb,amsfonts}
\usepackage{algorithmic}
\usepackage{graphicx}
\usepackage{textcomp}
\usepackage[table]{xcolor}
\usepackage{multirow}
\usepackage{booktabs, subcaption}
\usepackage{balance}

\begin{document}

\title{Traceability Transformed: Generating more Accurate Links with Pre-Trained BERT Models}

\def\posttrain{intermediate-train\ }
\def\posttrained{intermediate-trained\ }
\def\posttraining{intermediate-training\ }
\def\Posttraining{Intermediate-training }

\author{\IEEEauthorblockN{Jinfeng Lin, Yalin Liu, Qingkai Zeng, Meng Jiang, Jane Cleland-Huang}
\IEEEauthorblockA{\textit{Computer Science And Engineering} \\
\textit{University Of Notre Dame}\\
Notre Dame, IN, USA \\
jlin6, yliu26, qzeng, mjiang2, janehuang@nd.edu}
}

\newcommand{\todo}[1]{\textcolor{blue}{#1}}
\newcommand{\jlin}[1]{\textcolor{red}{#1}}
\newcommand{\yl}[1]{\textcolor{pink}{#1}}
\maketitle

\begin{abstract}
Software traceability establishes and leverages associations between diverse development artifacts. Researchers have proposed the use of deep learning trace models to link natural language artifacts, such as requirements and issue descriptions, to source code; however, their effectiveness has been restricted  by availability of labeled data and efficiency at runtime. In this study, we propose a novel framework called Trace BERT (T-BERT) to generate trace links between source code and natural language artifacts. To address data sparsity, we leverage a three-step training strategy to enable trace models to transfer knowledge from a closely related Software Engineering challenge, which has a rich dataset, to produce trace links with much higher accuracy than has previously been achieved. We then apply the T-BERT framework to recover links between issues and commits in Open Source Projects. We comparatively evaluated accuracy and efficiency of three BERT architectures. Results show that a Single-BERT architecture generated the most accurate links, while a Siamese-BERT architecture produced comparable results with significantly less execution time. Furthermore, by learning and transferring knowledge, all three models in the framework outperform classical IR trace models. On the three evaluated real-word OSS projects, the best T-BERT stably outperformed the VSM model with average improvements of 60.31\%  measured using Mean Average Precision (MAP). RNN severely underperformed on these projects due to insufficient training data, while T-BERT overcame this problem by using pretrained language models and transfer learning.

\end{abstract}


\section{Introduction}
\label{tb_sec:introduction}
Software and systems traceability, is the ability to create and maintain relations between software artifacts and to leverage the resulting network of links to support queries about the product and its development process. Traceability is deemed essential in safety-critical systems where it is prescribed by certifying bodies such as the USA Federal Aviation Administration (FAA), USA Food and Drug Administration (FAA)\cite{Rierson2013}. When present, trace links support diverse software engineering activities such as impact analysis, compliance validation, and safety assurance. Unfortunately, in practice, the cost and effort of manually creating and maintaining trace links can be inhibitive, and therefore trace links are typically incomplete and inaccurate \cite{mahmoud2012semantic}.  As a result, traceability data is often not trusted by developers and is often greatly underutilized.

Software artifacts, such as requirements, design definitions, code, and test cases all include natural language text, and therefore over the past decades, researchers have explored a wide variety of automated approaches for generating and evolving links automatically. Techniques have included probabilistic techniques \cite{Antoniol:Recovering}, the Vector Space Model (VSM),  \cite{DBLP:journals/tse/HayesDS06}, Latent Semantic Indexing \cite{DeLucia:ArtefManag,DBLP:conf/icse/RempelMK13}, Latent Dirichlet Allocation (LDA) \cite{DBLP:conf/re/DekhtyarHSHD07, DBLP:conf/icse/AsuncionAT10}, AI swarm techniques \cite{DBLP:journals/re/SultanovHK11}, recurrent neural networks  \cite{guo2017semantically} to integrate semantics, heuristic approaches \cite{DBLP:journals/jss/SpanoudakisZPK04, 6636704, DBLP:conf/re/Cleland-HuangMMA12}, combinations of techniques \cite{lohar2013improving, DBLP:conf/re/DekhtyarHSHD07, GethersICSM}, and the use of decision trees and support vector machines \cite{rath2018traceability} to integrate temporal dependencies and other process-related information into the tracing task. Despite all of these efforts, the accuracy of generated trace links has been unacceptably low, and therefore industry has been reticent to integrate automated tracing solutions into their development life-cycles. The primary impedance is a semantic one as most existing techniques rely upon word matching -- either direct matches (e.g., VSM), topic-based matches (e.g., using LSI or LDA), or indirect matches based on building a domain-specific ontology to bridge the terminology gap \cite{liu2020towards}. Results have been mixed, especially when applied to industrial-sized datasets, where acceptable recall levels above 90\% can often only be achieved at extremely low levels of precision \cite{DBLP:conf/sigsoft/LoharAZC13}.  

One of the primary reasons that automated approaches have underperformed is the semantic gap that often exists between related artifacts \cite{guo2017semantically}.  Techniques that are unable to reason about semantic associations and bridge this gap fail to establish accurate and relatively complete trace links. Recent work has proposed deep learning (DL) techniques \cite{borg2017traceability, zhao2017using} for traceability, but without providing effective solutions. For example, Guo et al. \cite{guo2017semantically}  proposed an architecture based on a Recurrent Neural Network (RNN) , and evaluated two types of RNN tracing models (LSTM\cite{hochreiter1997long} and GRU\cite{chung2014empirical}) for generating links between subsystem requirements and design definitions against a small dataset from an industrial project. While their results showed that accuracy improved as the size of the training set increased, their approach was not trained on large training sets, and therefore was not shown to generalize across larger or more diverse projects. We include both LSTM and GRU approaches for comparison purposes and refer to them collectively as TraceNN (TNN) in this paper.

Two primary factors impede the advancement of DL traceability solutions. The first is the sparsity of training data, given that DL techniques require large volumes of training data. Manually created trace links (i.e., golden answer sets) available in individual software projects are usually not sufficient for training a DL model. The second impedance is the practicality of applying  multi-layer neural networks in a large industrial project as training and utilizing deep neural networks is significantly slower than more traditional information retrieval or machine learning techniques.  

\begin{figure}[tbh]
        \centering
        \includegraphics[width=\columnwidth]{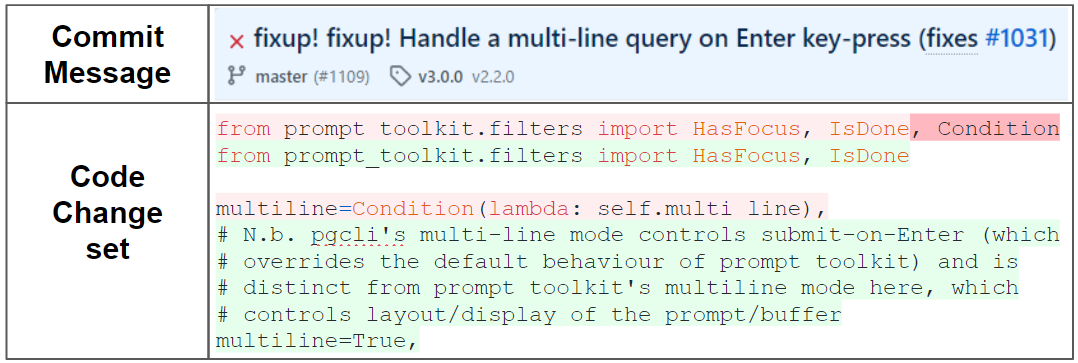}
        \hfill
        \includegraphics[width=\columnwidth]{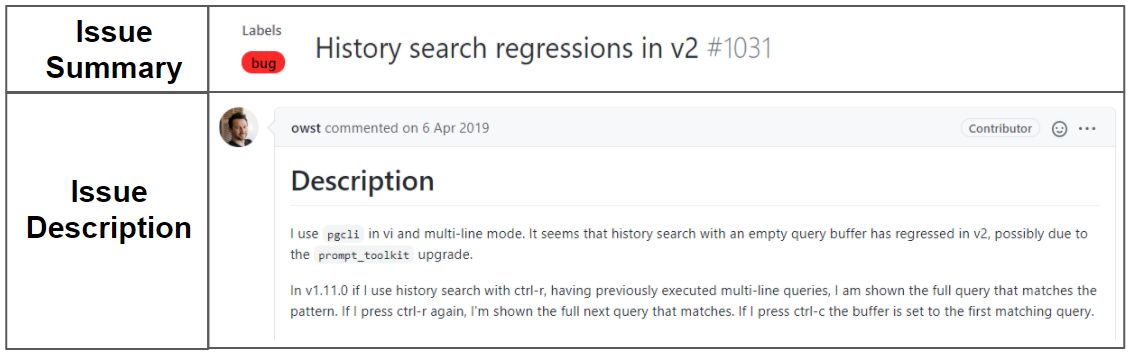}
    \caption{An example commit message and code change set, where green lines have been added and red ones removed.  The commit was tagged by the committer to the depicted issue.}
    \label{tb_fig:issue_commit_examples}
\end{figure}


The work reported in this paper addresses these two critical impedances in order to deliver fast and accurate automated traceability solutions for solving industrial problems.
More specifically, our proposed Language Model (LM) approach to traceability is designed to (1) deliver accurate and therefore  more trustworthy trace links, (2) be applicable for projects with limited training data, and (3) scale-up to support large industrial projects with low time complexity. Our approach leverages BERT (\textbf{B}idirectional \textbf{E}ncoder \textbf{R}epresentations from \textbf{T}ransformers) as its underlying language models. BERT, which was introduced by Google in 2018 \cite{devlin2018bert}, has delivered marked improvements in diverse NLP tasks, primarily because its bidirectional approach provides deeper contextual information than single-direction language models.  In this paper, we explore the use of BERT in the traceability domain -- introducing what we refer to as \emph{Trace BERT} (T-BERT). 

T-BERT is a framework for training a BERT-based relationship classifier to generate trace links. 
Three types of relation classification architectures are particularly well suited for traceability. These are the single, twin and, siamese  architectures, which we describe in more depth later in the paper.
We compare the effectiveness of these three architectures for generating trace links between Natural Language Artifacts (NLAs) and Programming Language Artifacts (PLAs). NLAs are artifacts such as feature requests, bug reports, requirements, and design definitions, which are written primarily using natural language but may also include code snippets. In contrast, PLAs are  primarily programming language artifacts, such as code files, code snippets, function definitions, and code change sets, which also contain natural language comments and descriptors.  We evaluated T-BERT by generating trace links from issues to code (represented by change sets), which we refer to as a \textbf{\textit{NLA-PLA traceability challenge}}.

The remainder of this paper is laid out as follows.  Sec.~\ref{tb_sec:problem_statement} outlines the concrete research questions we address in this paper. Sec.~\ref{tb_sec:approach} and Sec.~\ref{tb_sec:model_training} provide a detailed description of our approach for achieving NLA-PLA traceability, while Sec.~\ref{tb_sec:experiment} describes the experiments we conducted to evaluate the effectiveness of our approach.  Based on the results obtained from these experiments, we derive answers for our research questions in Sec.~\ref{tb_sec:discussion}.  Finally, Sec.~\ref{tb_sec:related_work} to Sec.~\ref{tb_sec:conclusion} discuss related work, threats to validity, and conclusions.

\section{Problem Statement}
\label{tb_sec:problem_statement}

Researchers have addressed the data sparsity problem and the performance issues of training large models through the use of pre-trained DL models for various NLP problems . This approach divides the training stage into pre-training and fine-tuning phases. In the pre-training phase, DL models are constructed using a huge amount of unlabeled data and self-supervised training tasks.
Then in the fine-tuning phase, the models are trained on smaller, labeled datasets in order to perform more specialized `downstream' tasks. The underlying notion is that knowledge learned from pre-training a model on a larger and more generalized dataset can be effectively transferred to the downstream tasks which have limited labels for supervised training. 
Furthermore, a pre-trained model provides a better starting point for model optimization than a randomly seeded one. It therefore reduces the likelihood of local optimization traps and improves overall performance. Fine-tuning a pre-trained model on a smaller dataset takes significantly less time than training a deep learning model from scratch. While pre-training a general model is extremely expensive, the pre-training phase only needs to be performed once and can then be reused for various downstream tasks. 


BERT-based language models make use of transformers \cite{vaswani2017attention} to learn contextual information from corpora in the pre-training stage and then transfer learned knowledge to downstream NLP tasks, such as question answering, document classification, and sentiment recognition  \cite{devlin2018bert, liu2019xqa}. To our knowledge, this is the first study that has applied BERT or other transformer-based methods to the software traceability task. 
We pose a series of research questions to evaluate whether T-BERT can effectively address the traceability problem. Our first question is defined as follows:  \vspace{3pt}\\
\textbf{RQ1}: Given three variants of T-BERT models, based on single, twin, and siamese BERT relation classifiers, which is the best architecture for addressing NLA-PLA traceability with respect to both accuracy and efficiency? \\ \vspace{-10pt}

In addition to investigating the DL model architecture, we also explore different training techniques for improving model accuracy. As discussed by  Guo et al. \cite{guo2017semantically}, the DL trace model may hit a `performance glass ceiling` and converge at relatively low accuracy. We therefore define our second research question as: \vspace{3pt}\\
\textbf{RQ2}: Which training technique improves accuracy without suffering from the previously observed glass ceiling ? \\ \vspace{-6pt}

Gururangan et al., in their study of Domain-Adaptive Pre-Training (DAPT), claim that a second phase of pre-training using a domain corpus leads to performance gains. This finding motivated us to explore the third and most important research question: \vspace{3pt}\\
\textbf{RQ3}: Can T-BERT transfer knowledge from a resource-rich retrieval task to enhance the accuracy and performance of the downstream NLA-PLA tracing challenge?\\ \vspace{-6pt}

Feng et al. \cite{feng2020codebert} demonstrated that a BERT Language Model, pre-trained using large numbers of function definitions, can effectively address the downstream code search problem. In that study, researchers provided doc-strings (i.e., python comments) as user queries, and leveraged a BERT model to retrieve related functions. Since doc-strings and functions are always paired in the code base, ample training data for the code search problem is available. Our RQ3 explores whether the code search problem can be leveraged as a training task to improve T-BERT for the software traceability challenge. Because this step occurs between pre-training and fine-tuning, we refer to it as \emph{\posttraining}

\section{Approach}
\label{tb_sec:approach}
Trace retrieval algorithms dynamically generate trace links between artifacts  \cite{cleland2014software}, for example, by linking a source artifact (e.g., a python file) to a target artifact (e.g., an issue or requirement).  The traceability algorithm computes the relevance between pairs of source and target artifacts and proposes the most related pairs as trace links. In this section, we first introduce the fundamental architecture of the BERT-based model and its variances, and then introduce T-BERT with three specific relation classifiers that are well suited for addressing this traceability problem. 

\subsection{Introduction to BERT and  Language Models}
A language model represents a probability distribution over a word sequence \cite{wiki:language_model}, which with proper training can effectively capture the semantics of individual words based on their surrounding context. Given the importance of general context, DL models built upon pre-trained language models usually achieve better results than those trained on task-specific datasets directly. 
The architecture of the BERT-based model is based on transformers, in which each layer in the model is a transformer layer. The transformer layers allow the BERT model to focus on terms at any position in a sentence, and training a BERT model is accomplished through a novel technique called Masked Language Modeling (MLM). In a MLM training task, BERT randomly masks the words in the input text and then optimizes itself to predict the masked terms based on the contextual information. In this pre-training step, a massive amount of corpora are fed to the BERT-based model, and the resulting model is leveraged to address different downstream tasks by fine-tuning on task-specific datasets. A distinctive feature of BERT is its unified architecture across different tasks \cite{devlin2018bert}, as the architecture for LM pre-training and task-specific fine-tuning are almost identical, with only the last layer of the model customized according to the targeted downstream tasks. This layer is usually referred to as a \textbf{\textit{task header}} in a BERT-based model.

\subsection{BERT For Software Traceability}
\label{tb_sub_section:bert_for_st}
\begin{figure}[t]
    \includegraphics[width=\linewidth]{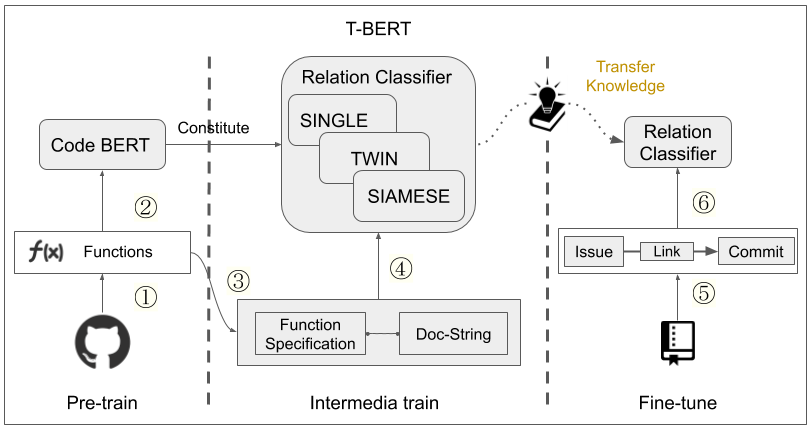}
    \caption{A three step workflow applies T-BERT to NLA-PLA traceability. 1) Pre-training data are functions collected from Github projects 2) A BERT is trained as a language model for code with these functions and composed with a relation classifier as the T-BERT model  3) Functions are split as specifications and doc-strings and used as \posttraining data 4) T-BERT model is \posttrained using code search data 5) OSS datasets are collected from Github repo 6) T-BERT model is fine-tuned as a trace model using transferred knowledge}
    \label{tb_fig:workflow}
\end{figure}

The solution we propose represents a three-fold procedure of pre-training, \posttraining and fine-tuning, as summarized in Fig.~\ref{tb_fig:workflow}.
In the pre-training phase, a dedicated language model is trained on source code and then utilized to construct the T-BERT models.
In the \posttraining phase, T-BERT is then trained to address the code search problem. In this phase, we provide adequate labeled training examples to T-BERT and expect it to learn general NL-PL classification knowledge that can ultimately be transferred to the  traceability challenge. 
Finally, in the fine-tuning phase, the \posttrained T-BERT model is applied to the issue-commit tracing challenge in real-world open-source projects.

\subsection{T-BERT Architectures}
\label{tb_sub_sec:tb_architecutre}
The three variants of the T-BERT architecture that we investigate for software traceability have previously been applied to similar text-based problems. These variants are:\vspace{3pt}

\noindent{$\bullet$ \bf TWIN:}
The Twin BERT architecture is shown in Fig.~\ref{tb_fig:twin_arch}. It leverages two BERT models to encode the NL and PL artifacts separately. The two artifacts are then transformed into two independent hidden state matrices, in which tokens are represented by fixed length vectors. We applied a pooling technique on these hidden state matrices to formulate feature vectors representing the artifacts. Finally, we concatenated these two feature vectors for classification tasks.\vspace{3pt}



\noindent{$\bullet$ \bf SIAMESE:}
The siamese BERT architecture is shown in Fig.~\ref{tb_fig:siamese_arch}. It is a hybrid of the single and twin architecture. It only uses one BERT model; however, instead of creating a concatenated token sequence for an NL-PL pair like single BERT, it passes each artifact sequentially to the BERT model and creates separate hidden state matrices for each of the two artifact types (i.e., NL and PL). The generated two hidden state matrices are then pooled and concatenated to produce a joint feature vector as in the Twin BERT architecture. This joint feature vector is then sent to classification headers to accomplish the prediction task. 

Both Siamese and Twin T-BERT architectures concatenate artifact feature vectors to create a joint feature vector. Nils et al. \cite{reimers2019sentence} explored the impact of different concatenation approaches for the  siamese BERT architecture and showed that given two pooled feature vectors $u$ and $v$, siamese BERT with a joint feature vector ($u,v, |u-v|$) achieved the best performance on a sentence classification task. We therefore apply this type of concatenation method to fuse the NL and PL feature vectors to create a joint feature vector. \vspace{3pt}

\noindent{$\bullet$ \bf SINGLE:}
The single BERT architecture is shown in Fig.~\ref{tb_fig:single_arch}. NL and PL text are annotated with special tokens and then concatenated into a single sequence. For example, token [CLS]/[SEP] is used to annotate the start/end of a sentence. A sentence S with tokens $s_1, s_2, s_3,.. s_N$, and a piece of code C with tokens $c_1, c_2, c_3,.. c_N$ will be transformed into an input format of $[CLS] s_1, s_2, s_3,.. s_N, [SEP] c_1, c_2, c_3,.. c_N [SEP]$. The annotated and concatenated sequence is fed to the single BERT to generate a single hidden state matrix. A subsequent pooling layer then reduces the dimension of the matrix to create a fused feature vector, which is a counterpart of the joint feature vector in SIAMESE and TWIN. This feature vector is used by the classification header, to predict whether the input NL-PL pair is related or not. 

\section{Model Training}
\label{tb_sec:model_training}
In this section, we  describe the training strategies used for pre-training, \posttraining, and fine-tuning phases. The dataset supporting pre-training and \posttraining phases is provided by Hamel et al. \cite{husain_codesearchnet_2019} from their study of the code search problem. It includes function definitions and their associated doc-strings scraped from numerous Github projects, and includes Go, Java, JavaScript, PHP, Python, and Ruby programming languages.

The dataset used in the fine-tuning phase was retrieved from OSS by our team.  We extracted issues and commits through Github's APIs and mined ground-truth trace links from the commit messages. We show the data format in Fig.~\ref{tb_fig:issue_commit_examples}, and explain details of the data collection process in Sec.~\ref{tb_sub_sec:data_collection}. 
In this study we selected Python as our target language for both training and evaluation due to the large number of active projects; however, our approach is not language dependent. Given sufficient time, the same post-training process could be applied to other programming languages. 

\subsection{Three Step Training}

\noindent{$\bullet$ \bf Pre-training Code Language Model:} In the pre-training step, we leveraged the BERT model to learn the word distribution among NL and PL documents, and refer to this BERT model as `code BERT' to distinguish it from the plain BERT model that handles only NL text. In the plain BERT model, masked LM (MLM) tasks were used to pre-train BERT as a language model. As previously explained, in MLM tasks, 15\% of tokens are selected and masked, and then BERT is trained to recover the masked tokens based on their surrounding context. 

\begin{figure}[t!]
    \begin{subfigure}{1\columnwidth}
        \centering
        \includegraphics[width=.9\columnwidth]{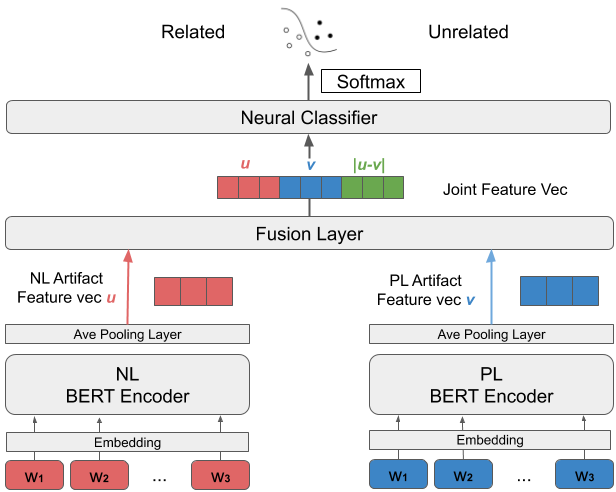}
        \caption{TWIN}\label{tb_fig:twin_arch}
        \vspace{12pt}
    \end{subfigure}

    \begin{subfigure}{1\columnwidth}
        \centering
        \includegraphics[width=.9\columnwidth]{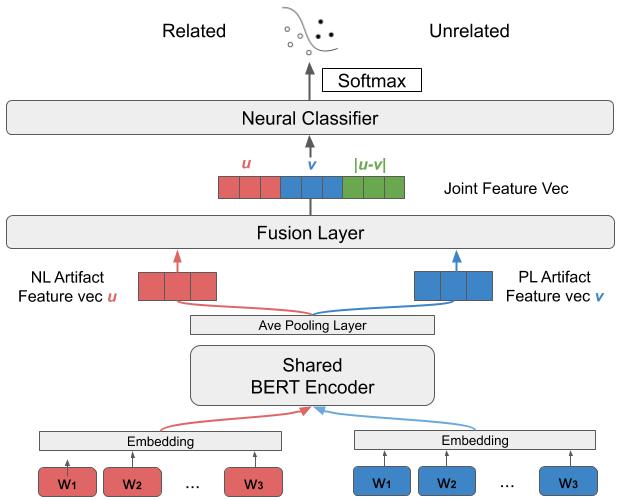}
        \caption{SIAMESE}\label{tb_fig:siamese_arch}
        \vspace{12pt}
    \end{subfigure}
    \vspace{6pt}
    \begin{subfigure}{1\columnwidth}
        \centering
        \includegraphics[width=.9\columnwidth]{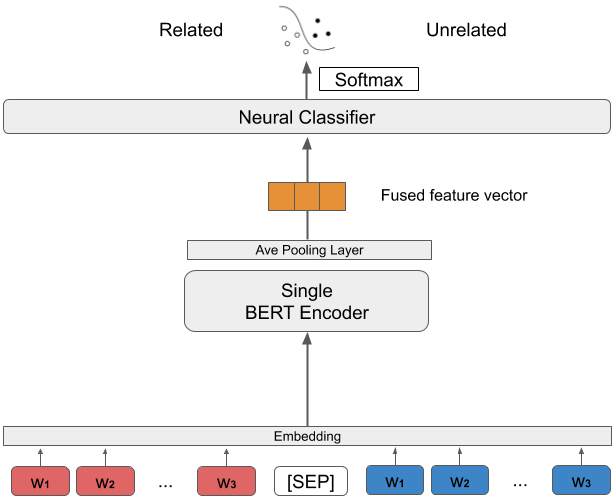}
        \caption{SINGLE}\label{tb_fig:single_arch}
    \end{subfigure}
    \caption{The architectures of the three T-BERT models proposed and evaluated in our experiments.}
    \label{tb_fig:tb_arch}
    \vspace{-12pt}
\end{figure}

Given that pre-training a language model is very expensive, three commercial organizations have released their own pre-trained code BERT models (Hugging Face \cite{hg_web}, CodistAI \cite{codistAI_web}, and Microsoft \cite{feng2020codebert}), all of which were trained on the CodeSearchNet dataset. Of these, we leverage Microsoft's model (referred to as MS-CodeBert) as our source code language model directly for T-BERT relation classification models depicted in Fig.~\ref{tb_fig:tb_arch}, as it has been shown to deliver improved language comprehension for diverse downstream software engineering tasks. These improvements in MS-CodeBert can be attributed to its `Replaced Token Detection' training tasks, which have been shown to be a more effective way of training LMs \cite{clark2020electra}. This training task replaces a small portion tokens in the corpus with random tokens and then requires the BERT model to identify which tokens have been replaced in the corpus. 


\noindent{$\bullet$ \bf Intermediate-training:} For \posttraining we trained T-BERT models to perform the code search problem, as this problem is inherently similar to the NLA-PLA traceability challenge. In both cases, we used T-BERT to retrieve related source code based on a NL description of code functionality. 
The CodeSearchNet dataset \footnote{CodeSearchNet dataset https://github.com/github/CodeSearchNet} provides a benchmark for the code search problem, as each function in the dataset is paired with a doc-string. For Python, the dataset includes
824,342 functions for training, 46,213 functions for development and 22,176 functions for testing. This dataset is ideal for \posttraining purposes because 1) it is large in size, therefore the T-BERT model has adequate labeled data to learn general rules for identifying NL-PL relevance, 2) the relationships between doc-string and function are definitive, meaning that there is minimal noise in the ground truth, and 3) the function definitions use only part of the python grammar which makes this task easier to handle than NLA-PLA traceability. 

We formulated the \posttraining as a binary classification task in which T-BERT was asked to identify whether a given doc string properly describes its paired function or not. The loss function used in this \posttraining step was Cross Entropy Loss, and Adam Optimizer\cite{kingma2014adam} was used to update the parameters and optimize for the loss function.

The code search problem creates an unbalanced distribution of positive and negative docstring-to-code pairs, we therefore created a balanced training dataset with an equal number of positive and negative samples. Guo et al. \cite{guo2017semantically} adopted a dynamic under-sampling strategy to avoid inflating training data size while continually exposing the model to previously unseen negative samples. We adopted a similar technique to construct our training samples. In each epoch, a balanced training set was constructed by including all function and doc-string pairs from CodeSearchNet dataset, as well as a randomly selected equal number of non-related pairs. We updated the training set at the beginning of each epoch, so that the T-BERT model could learn from previously unseen negative examples. We refer to this training strategy as Dynamic Random Negative Sampling (DRNS), and compare it to other training strategies described in Sec.~\ref{tb_sub_sec:neg_sampling}.

\noindent{$\bullet$ \bf Fine-tuning:}  In fine-tuning, we utilized a similar training technique to that discussed in the previous step, but addressed the traceability challenge of tracing issues to code commits using real world OSS datasets. Although the input data is formatted differently to the \posttraining format, T-BERT uses the same architecture for both tasks. As shown in Fig.~\ref{tb_fig:issue_commit_examples}, the issues are comprised of a short issue summary and a long issue description while the commit is composed of a commit message and code change set. For each type of artifact, we concatenated the text to formulate input sequences for the T-BERT model (i.e., issue summary + issue description and then commit message + code change set.) In contrast to the \posttraining step, the dataset utilized in this step is limited and fuzzy. As reported by Rath et al., link sets mined from OSS projects are unlikely to be complete and entirely accurate as engineers may forget to tag issues, or may commit multiple changes against multiple tags. \cite{rath2018traceability}. Furthermore the number of links in the project-specific fine-tuning phase is significantly smaller than the number of function to doc-string pairs used in post-training. As reported in Table.~\ref{tb_tab:oss_data}, we have approximately six hundred true links for fine-tuning compared to 824,342 pairs of function and doc-string. Furthermore, the code change set in commits has a more flexible and complex format than the short and succinct function definitions.

\subsection{Negative Sampling}
\label{tb_sub_sec:neg_sampling}
Guo et al., observed a glass ceiling in terms of achieved trace link performance in which the accuracy of their neural trace model increased as the training epoch increased at the beginning, however, it then reached a peak value and started to decline with further training \cite{guo2017semantically}. Our hypothesis of this phenomenon is as follows. At earlier stages of training, the trace model can effectively learn rules for distinguishing positive and negative examples, and the neural trace model was easily able to rule out many unrelated examples (i.e.,  cases where the source and target artifacts have little common vocabulary). Since those types of negative examples constitute the majority of negative examples, the random negative sampling strategy experiences few challenging examples and therefore starts overfitting based on n\"aive rules, because these n\"aive rules are applicable for the majority of simple cases. This overfitting causes accuracy to decline after a few training epochs.

Our approach adds high quality negative samples to alleviate this problem through our proposed Online Negative Sampling (ONS) as an alternative to Dynamic Random Negative Sampling (DRNS). The principle idea of ONS is that, instead of creating a training dataset at the beginning of each epoch, the trace model will generate negative examples dynamically at the batch level. 
For illustrative purposes imagine we want to create a batch of size 8 containing 4 positive links. If we have 4 NL artifacts and 4 PL artifacts (i.e., a total of 16 candidate links), we include the 4 positive links, and then select 4 negative links from the 12 candidates in order to create a balanced batch. By evaluating these negative links and ranking them by predicted score, we can identify the false links which are more likely to be mistakenly classified. Then by incorporating the top-scoring negative examples into the training set, we improve the quality of negative examples and avoid early overfitting. This approach is inspired by applications in the Face Recognition domain, where face recognition models need to distinguish between people with similar appearances \cite{schroff2015facenet}. This negative example mining strategy is usually combined with Triplet Loss \cite{wiki:trip_loss} in the contrastive learning \cite{chen2020simple} framework. Here we adopt it for our  classification and combine it with the widely used Cross Entropy Loss.

\section{Experimental Evaluation }
\label{tb_sec:experiment}
\subsection{Data Collection}
\label{tb_sub_sec:data_collection}
In this study, we train T-BERT and evaluate it against two types of datasets. The first dataset is CodeSearchNet which is publicly available  \cite{husain_codesearchnet_2019}. It  includes functions and their associated doc-strings for six different programming languages. As previously stated, we focused on python functions.

The other datasets\footnote{ OSS dataset https://zenodo.org/record/4511291\#.YB3tjyj0mbg} we leveraged were mined from three OSS projects in Github and included Pgcli, Flask and Keras as described in Table.~\ref{tb_tab:oss_data}. We selected these projects because they are popular, actively maintained Python projects, with developers actively tagging commits with issue IDs.
We retrieved issues and their discussions as the source artifacts and commits as target artifacts. For each issue, we included both the short issue summary and the longer issue description. We automatically removed stack traces from issue discussions if highlighted as Code block in markdown, as we wanted to train our approach to perform the harder job of generating links between issues and code without the more explicit information provided by stack traces. 
For each commit, we included the commit message and change set. However, we removed very small commits (with less than 5 LOC) from our target artifact set.  Finally, due to the Github API rate limit, we retrieved a maximum of 5000 issues for each project. 

After retrieving commits and issues we mined a `golden link set' from the commit messages by using issue tags embedded into commit messages added by committers. In addition, we leveraged pull requests, as an accepted pull request automatically creates both an issue and commit in the OSS project, and connects them through an issue ID embedded into the commit message. Tags were mined using regular expressions in order to build a link set connecting issues and commits. One risk of mining links from commit message is that the link set may be incomplete. Liu et al. partially addressed this problem by pruning the dataset and only retaining artifacts appearing in links set \cite{liu2020traceability}. We adopted this process to construct our dataset and report results in  Table~\ref{tb_tab:oss_data}

\begin{table}[tbh!]
	\centering
    \caption{The size of software project leveraged in traceability experiment. We applied the cleaning procedures, described in Sec.~\ref{tb_sub_sec:data_collection}, to clean artifacts and links.  }
    
    \begin{tabular}{cccccc}
    \toprule \hline 
    & & Commits & Issues & Links & Type \\ \hline
    \multirow{2}{*}{ Pgcli } & original & 2191 & 1197 & 645 & \multirow{2}{2cm}{\centering database command line } \\
    & cleaned & 531 & 522 & 530 & \\ \hline
    \multirow{2}{*}{ Flask } & original & 4011 & 3711 & 1159 & \multirow{2}{2cm}{\centering web framework }\\
    & cleaned & 752 & 739 & 753 & \\ \hline
    \multirow{2}{*}{ Keras } & original & 5348 & 4810 & 9375 & \multirow{2}{2cm}{\centering neural-network library } \\
    & cleaned & 551 & 550 & 551 & \\ \hline
    \end{tabular}

    \label{tb_tab:oss_data}
\end{table}

\subsection{Experiment Setup}
We conducted our experiment on Supermicro SYS-7048GR-TR SuperServer with Dual Twelve-core 2.2GHz Intel Xeon processors and 128GB RAM. We utilized 1 NVIDIA GeForce GTX 1080 Ti GPU with 10 GB memory to train and evaluate our model. The T-BERT model was implemented with PyTorch V.1.6.0 and HuggingFace Transformer library V.2.8.0. 
We trained models for 8 and 400 epoch in \posttraining and fine-tuning. For each task, we use a batch size of 8 and a batch accumulation step of 8. We set the initial learning rate as 1E-05 and applied a linear scheduler to control the learning rate at run time. 
Regarding the model selection, we split the dataset into training (train), development (dev), and test sets. We trained the model using the training dataset and tested its performance on the dev dataset. We then selected the best performing model based on the dev dataset and created an output model. We finally evaluated and compared the performance of output models on the test dataset. In the  \posttraining stage the dataset was already split by the data provider. In the fine-tuning stage, we split the dataset into ten folds, of which eight were used for training, one for development, and one for testing.

\subsection{Evaluation Metrics}
The metrics for our experiments include F-scores, Mean Average Precision (MAP@3), Mean Reciprocal Rank (MRR), and Precision@K. \vspace{3pt}

\noindent{$\bullet$ \bf F-scores:} F-scores are composite metrics calculated from precision and recall, and are frequently used to evaluate traceability results. The F-1 score assigns equal weights to precision and recall, while the F-2 score favors recall over precision. Although both precision and recall are important, F-2 is usually preferred for evaluating trace results where recall is considered more important than precision. We report the best F-scores in our experiments by enumerating the thresholds. 

\begin{equation}
F_{\beta}= \frac{(1+\beta^2) \cdot precision \cdot recall}{\beta^2 \cdot precision \cdot recall }
\end{equation}\vspace{3pt}

\noindent{$\bullet$ \bf Mean Average Precision:} MAP evaluates the ranking of relevant artifacts over retrieved ones. Each source artifact is regarded as a query Q for retrieving artifacts. After ranking the retrieved target artifacts, an Average Precision (AveP or AP) score is obtained based on the position of all relevant target artifacts in the ranking. The Mean of AveP scores is then computed to return the MAP. In our study, we apply a stricter metric known as MAP@3, in which only artifacts ranked in the top 3 positions contribute to $AveP$ score. The formula for this metric is shown in following, k represents the total relevant target artifacts in a query and $rank_i$ refer to the ranking a target artifact :\\
\begin{equation}
    AveP@3 = \frac{\sum_{i}^{k} X}{k}, \:    X=
    \begin{cases}
      P@i, & \text{if}\ rank_i<=3 \\
      0, & \text{otherwise}
    \end{cases}
\end{equation}
\begin{equation}
MAP@3 = \frac{\sum_{j}^{Q} AveP_j@3}{Q}
\end{equation}

\noindent{$\bullet$ \bf Mean Reciprocal Rank:} MRR is another measurement of the result ranking. In each query, the first related target artifact with a rank of N, will provide a Reciprocal Rank of 1/N. MRR accumulates by averaging the Reciprocal Rank for all the queries Q. This focuses on the first effective results for a query, while ignoring the overall ranking. This is the standard metric used for the CodeSearchNet benchmark. While MAP is more typical for trace retrieval tasks, we include this metric to compare our \posttrained model against the approaches in other studies for code search problem.  
\begin{equation}
MRR= \frac{1}{Q}\sum_{|i=1|}^{Q} \frac{1}{1stRank_i}
\end{equation}

\noindent{$\bullet$ \bf Precision@K:} Precision@K evaluates how many related artifacts are retrieved and ranked in the top K. The formula for this metric is shown in Eq.~\ref{tb_eq:p_at_k}. We provide results with K values of 1 to 3. A trace model with high Precision@K means users are more likely to find at least one related target artifact in the top K results. 
\begin{equation}
\label{tb_eq:p_at_k}
Precision@K = \frac{\sum_{i}^{Q}|Rel_i@K|}{|Rel|}
\end{equation}

As we can see, MRR and Precision@K ignore recall and focus on evaluating whether the search result can find interesting results for a user. They are ideal for the code search problem but not for traceability where recall is particularly important. Therefore, we apply only F-score and MAP@3 to evaluate traceability results.As the majority of our queries have fewer than three correct links, a perfect MAP@3 score represents close to 100\% recall.
\section{Results And Discussion}
\label{tb_sec:discussion}
We report performance results of T-BERT models for the code search problem and the NLAs-PLAs traceability problem, and address the RQs defined in Sec.~\ref{tb_sec:introduction}.

\subsection{Evaluating the Code Search Problem}
Our first evaluation explores how well T-BERT models perform when datasets have adequate labeled examples.
We trained T-BERT models for the three architectures introduced in Sec.~\ref{tb_sub_sec:tb_architecutre}, using the training part of the CodeSearchNet dataset. For the T-BERT models which are trained with ONS technique, we add a star sign after the model names to distinguish them from the T-
BERT model trained with DRNS. For example, SINGLE* refers to the model with single BERT architecture and trained with online negative sampling. The performance of these six types of models are reported in Table.~\ref{tb_tab:csn_acc}. In addition, we compare T-BERT models against three classical tracing techniques of VSM, LDA, LSI and also a deep learning trace model, TraceNN \cite{guo2017semantically} for this dataset. 

\begin{figure*}[thb!]

    \label{fig: exp1_bar}
    \begin{subfigure}{.32\linewidth}
        \centering
        \includegraphics[width=\linewidth]{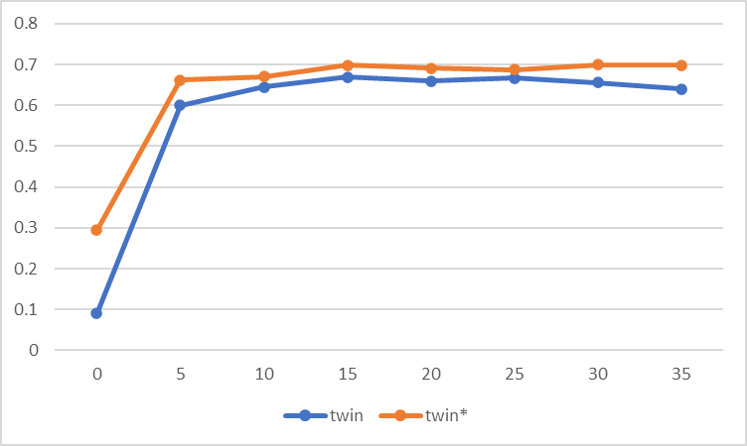}
        \caption{TWIN}\label{tb_fig:twin_curv}
    \end{subfigure}
        \hfill
    \begin{subfigure}{.32\linewidth}
        \centering
        \includegraphics[width=\linewidth]{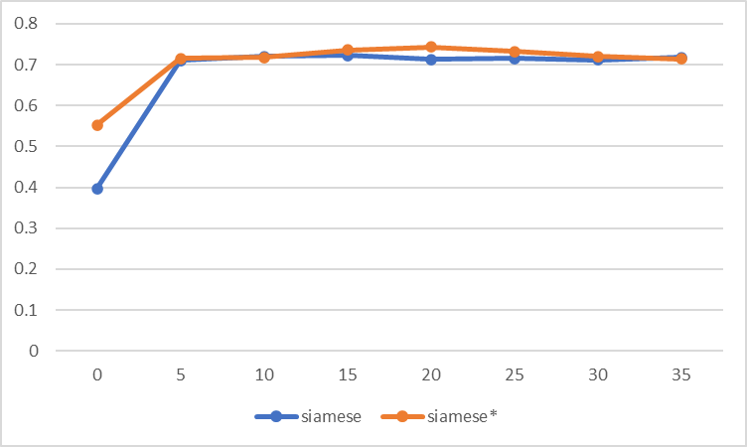}
        \caption{SIAMESE}\label{tb_fig:siamese_curv}
    \end{subfigure}
       \hfill
    \begin{subfigure}{.32\linewidth}
        \centering
        \includegraphics[width=\linewidth]{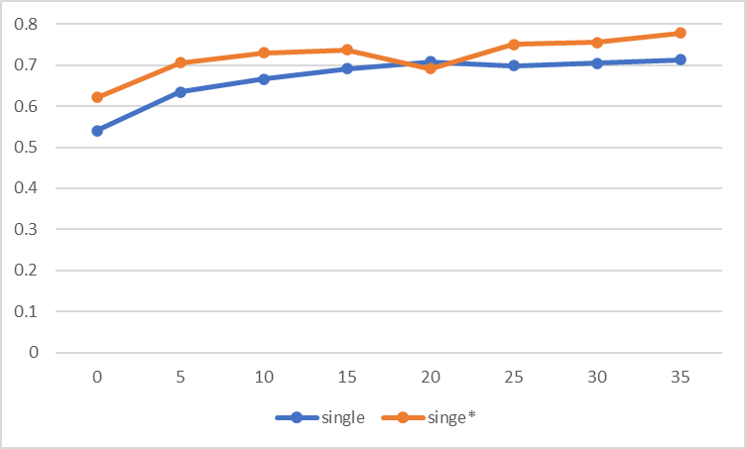}
        \caption{SINGLE}\label{tb_fig:single_curv}
    \end{subfigure}
    \caption{The learning curve for T-BERT and T-BERT* models on code search challenge. This figure shows the MAP scores (Y-Axis) over the first 35K (X-Axis) Adam optimization steps.}
    \label{tb_fig:learn_curve}
\end{figure*}

Other researchers such as Hamel et al. \cite{husain_codesearchnet_2019} and Feng \cite{feng2020codebert} leveraged the same dataset to conduct code search study, and so we select the methods which achieved the best MRR scores in their study as a comparison to T-BERT, and created our evaluation dataset for CodeSearchNet in the same way. For each doc-string, we combined the related function with 999 unrelated ones, and charged the retrieval models with finding the correct function among one thousand candidates. However, the SINGLE models were not able to efficiently process the entire dataset, and so in this case we evaluated only 100 out of the total 22,176 queries.  The comparison of MRR scores are shown in Table.~\ref{tb_tab:csn_acc}.
\definecolor{col}{HTML}{ ecf0f1 }
\begin{table}[tbh]
	\centering
	\caption{Evaluation of T-BERT models on the CodeSearchNet Challenge dataset}
	\setlength\tabcolsep{4.2pt}
        \begin{tabular}{l|cccrcccc}
        \toprule \hline
            & F1 & F2 & MAP & MRR & Pr@1 & Pr@2 & Pr@3 \\
            TWIN & 0.497 & 0.563 & 0.735 & 0.756 & 0.646 & 0.787 & 0.842 \\
            SIAMESE & 0.604 & 0.668 & 0.814 & 0.825 & 0.729 & 0.866 & 0.915 \\
            SINGLE &  0.482& 0.572 & 0.825 & 0.839 & 0.730 & 0.900 & 0.930\\
            \hline
            TWIN* & 0.559 & 0.626 & 0.794 & 0.809 & 0.712 & 0.846 & 0.890 \\
            SIAMESE* & 0.594 & 0.655 & 0.817 & 0.829 & 0.738 & 0.866 & 0.910 \\
            SINGLE* & 0.612 & 0.678 & 0.837 & 0.851 & 0.750 & 0.910 & 0.930 \\ 
            \hline
            VSM & 0.219 & 0.255 & 0.314 & 0.351 & 0.251 & 0.341 & 0.397 \\
            LDA & 0.005 & 0.010 & 0.012 & 0.021 & 0.008 & 0.013 & 0..017 \\
            LSI & 0.003 & 0.007 & 0.014 & 0.025 & 0.009 & 0.015 & 0.020 \\ \hline
            TNN-LSTM  & 0.179 & 0.245 & 0.351 & 0.400 & 0.269 & 0.386 & 0.457  \\
            TNN-BiGRU & 0.221 & 0.29  & 0.392 & 0.438 & 0.304 & 0.432 & 0.504   \\
            JV-biRNN   &\cellcolor{col} &\cellcolor{col} &\cellcolor{col} & *0.321 &\cellcolor{col} &\cellcolor{col} &\cellcolor{col} \\
            JV-SelfAtt & \cellcolor{col}& \cellcolor{col}&\cellcolor{col} & *0.692 &\cellcolor{col} &\cellcolor{col} &\cellcolor{col} \\
            MSC & \cellcolor{col}& \cellcolor{col}& \cellcolor{col}& *0.860 &\cellcolor{col} &\cellcolor{col} & \cellcolor{col}\\
            \hline \bottomrule 
        \end{tabular}\\
        \vspace{3pt}
        JV=Joint Vector \cite{husain_codesearchnet_2019}; MSC=MS-CodeBERT \cite{feng2020codebert}; TNN = TraceNN\cite{guo2017semantically}; \\
        *Previously reported results against same CodeSearch-Net challenge dataset.
	\label{tb_tab:csn_acc}
\end{table}
\begin{table}[hb]
	\centering
    \caption{Training and testing time for T-BERT models on code search problem. The test time is recorded for a test set with 100 queries.}

        \begin{tabular}{ccccc} 
            \toprule
            &Strategy& TWIN & SIAMESE & SINGLE \\ \hline
            Train(hr)&DRNS  & 156h & 138h & 164h \\
            Test(sec)&DRNS & 3254s & 3264s & 183357s \\
            \midrule
            Train(hr) &ONS& 146h & 142h & 283h \\
            Test(sec)&ONS & 3211s & 3265s & 193667s \\ 
            \bottomrule
        \end{tabular}
        \label{tab:csn_time}
\end{table}
To observe the learning process for each model, we visualized the learning curve in Fig~.\ref{tb_fig:learn_curve} for the first 35,000 steps of optimizations. We evaluate the performance of each model at intervals of 1000 steps during the training by applying the intermediate model against small testing sets composed of 200 development examples.

\subsection{Evaluate NLA-PLA Traceability}
We then evaluated the T-BERT models on the NLA-PLA Traceability problem. As previously described, we used 8 folds of trace links for training, 1 fold for developing and 1 fold for testing. To explore our RQ3, we conducted a controlled experiment, in which we trained two groups of T-BERT models. In the first group, we continued training the T-BERT model which had been \posttrained in our previous experiment. In the second group, we trained the T-BERT models without applying transferred knowledge learned from \posttraining. When we conducted model training, we applied the same training dataset and ONS techniques to the two groups. To maintain consistency of abbreviations, we name the models in the first group, for example, as SINGLE*+T; and the models in the second group as SINGLE*. The result of this experiment are shown in Table.~\ref{tb_tab:trace_acc}, while the learning curve for the T-BERT models showing the fine-tuning to traceability tasks is shown in Fig.~\ref{tb_fig:trace_curve}. We show only the learning curve for Pgcli data due to space constraints.


\begin{table*}[tbh!]
	\centering
	\caption{Evaluation of models on NLAs-PLAs traceability}
	\addtolength{\tabcolsep}{+2pt}
	    \begin{tabular}{l|c|ccc|ccc|ccc}
	    \toprule \hline
        & & \multicolumn{3}{c|}{ Pgcli } & \multicolumn{3}{c|}{ Flask } & \multicolumn{3}{c}{Keras} \\ 
        & & F1 & F2 & MAP & F1 & F2 & MAP & F1 & F2 & MAP \\  \hline
        TWIN* & & 0.450 & 0.491 & 0.574 & 0.524 & 0.577 & 0.683 & 0.450 & 0.491 & 0.574 \\
        SIAMESE* & & 0.621 & 0.654 & 0.728 & 0.681 & 0.731 & 0.801 & 0.962 & 0.962 & 0.990 \\
        SINGLE* & & 0.707 & 0.745 & 0.785 & 0.841 & 0.873 & 0.952 & 0.931& 0.925 & 0.971 \\ \hline
        TWIN* & T & 0.686 & 0.709 & 0.766 & 0.750 & 0.781 & 0.869 & 0.953 & 0.970 & 0.978 \\
        SIAMESE* & T & 0.729 & 0.748 & 0.779 & 0.820 & 0.830 & 0.920 & 0.971 & 0.977 & 0.990 \\
        SINGLE* & T & 0.730 & 0.789 & 0.859 & 0.884 & 0.862 & 0.92 & 0.972 & 0.989 & 0.990 \\ \hline
        VSM & & 0.376 & 0.424 & 0.506 & 0.509 & 0.474 & 0.540 & 0.532 & 0.512 & 0.703 \\
        LDA & & 0.121 & 0.226 & 0.208 & 0.182 & 0.241 & 0.227 & 0.290 & 0.367 & 0.333 \\
        LSI & & 0.085 & 0.145 & 0.147 & 0.127 & 0.164 & 0.142 & 0.072 & 0.126 & 0.109 \\
        TNN-LSTM  & & 0.138 &0.179 & 0.128 & 0.106 & 0.126 & 0.080 & 0.053 & 0.087 & 0.034\\ 
        TNN-BiGRU & & 0.062 &0.116 & 0.006 & 0.066 & 0.100   & 0.044 & 0.063 & 0.119 & 0.073 \\ 
        \hline
        \bottomrule
        \end{tabular}\\
        \vspace{3pt}
        T=Transfer learning, TNN = TraceNN\cite{guo2017semantically}; 
	    
        
        

	\label{tb_tab:trace_acc}
\end{table*}
\begin{table}[hb]
	\centering
	\caption{Model performance on Pgcli dataset for NLA-PLA.}
	\begin{tabular}{cccc}\toprule
        & TWIN* & SIAMESE* & SINGLE* \\ \hline
        Train(hr) & 12h & 12h & 13h \\
        Test(sec) & 170s & 163s & 5395s \\
        
        \bottomrule
        \end{tabular}
	\label{tb_tab:trace_time}
\end{table}

\subsection{RQ2: How does ONS alleviate the  glass ceiling problem?}
In this section, we discuss how the effectiveness of ONS for T-BERT models alleviates the glass ceiling problem. This question helps us to identify the best approach for training T-BERT models. To answer this question we apply both ONS and DRNS to the same T-BERT architecture to create test and control groups.  Fig.~\ref{tb_fig:learn_curve} shows the learning curve of T-BERT models on CodeSearchNet dataset during training. The orange lines represent the models trained with ONS while blue lines represent those trained with DRNS. We find that for TWIN and SINGLE model, the orange line is always above the blue line, meaning that ONS can accelerate the learning for T-BERT models but also let it converge at a higher value. While for SIAMESE we find the orange line is above blue line in the early steps, but soon converges at a similar level. This result indicates ONS benefits the T-BERT model training by introducing harder negative examples. 
The evaluation results shown in the first six rows of Table.~\ref{tb_tab:csn_acc} also support this finding, as the TWIN* and SINGLE* models (line 4,6) achieve better results than T-BERT models (line 1,3) from the perspectives of all metrics. SIAMESE* (line 5) and SIAMESE (line 2) have a very close result where the difference of all metrics are within 0.5\% except F2 score (1.3\%).

We report the training time for DRNS and ONS in Table~\ref{tab:csn_time}.  ONS introduces initial overheads in constructing each batch, but then has fewer candidates to evaluate and sort.  In contrast, DRNS has no upfront construction costs, but must sample data from a large list, creating a performance bottleneck. The use of ONS only significantly increased training time except for the SINGLE model which is particularly slow on evaluation. 

We conclude that ONS delivers better accuracy than DRNS, and only increases training times for models (e.g., SINGLE) which have slow evaluation processes.


\subsection{RQ1: Which T-BERT architecture is better?}
This RQ focuses on comparing the accuracy and efficiency of T-BERT when used for trace retrieval tasks. Performance comparisons were conducted on T-BERT* models, as we have shown in RQ2 that T-BERT* models returned better accuracy than T-BERT ones. To answer this question, we further divided our questions into three sub RQs as following. 

\noindent{$\bullet$ \bf RQ1.1:} Are T-BERT models capable of resolving the CodeSearchNet and NLA-PLA traceability problem? 

Table.~\ref{tb_tab:csn_acc} shows that SINGLE*, TWIN*, and SIAMESE* (line 4-6) can achieve F scores around 0.6 and MAP scores around 0.8. The Precision@3 score for the three models are around 0.9 which means T-BERT* models can return related functions in around 9 of 10 user queries. And in  75\% to 80\%  of cases the correct answer definition is ranked at the first position. This result shows that all three models are effective for the CodeSearchNet challenge. Among these three models, SINGLE* achieves the best performance with respect to all metrics. However, the gap between these three models is small, and all three models clearly outperform the base line created by the three IR models of VSM, LSI, and LDA. 

In Table.~\ref{tb_tab:trace_acc}, the first three rows show that T-BERT* models applied to the traceability challenge and trained without the benefit of transfer knowledge are ranked in the same way as for the code search problem. However, the performance gap between these three models increases. This suggests that the size of training data  has different impacts on the three types of architecture. Since, TWIN* model includes two inner BERT models, the parameters in this architecture are doubled, and it therefore requires more training examples to tune the models. Nevertheless, all T-BERT* models achieved achieve better results than the IR model baselines. Especially in Keras dataset, SIAMESE* and SINGLE* (line2 and 3) have an F score above 0.95 and MAP of 0.99 indicating that T-BERT can provide perfect tracing results in some scenarios.

\noindent{$\bullet$ \bf RQ1.2:} Which T-BERT model most effectively addresses the two problems of data sparsity and performance ?

We need to take both accuracy and efficiency into consideration when selecting a model for use in production.  As discussed in RQ1.1, the SINGLE* model achieves best performance; however, it is very slow for processing large scale datasets. As shown in Table.~\ref{tab:csn_time}, TWIN and SIAMESE need around 3000 seconds to evaluate 100 queries while SINGLE needs around 20000 seconds. We estimate that it will take around 6000 hours for SINGLE to evaluate the whole CodeSearchNet test with our current experiment setup. But for TWIN and SIAMESE, it took us only around 20 hours to evaluate the whole test set in practise. 
In the traceability challenge, the test set is relatively tiny. Taking Pgcli for example, it contains 2704 candidate links compose of 52 source artifact and 52 target artifacts. TWIN and SIAMESE both take around 160 seconds to finish the task while the SINGLE model takes around one hour.
SIAMESE and TWIN architectures accelerate the process by decoupling the feature vector creation steps. In the SINGLE architecture, NL and PL document pairs are fed to BERT to create a joint feature vectors. This step is extremely expensive and creates the main performance bottleneck for the SINGLE model. Assuming we have N source and N target artifacts. SINGLE has a time complexity of $O(N^2*K)$ for creating feature vectors for all the candidate links, where K refer to the time consumed by the BERT model to convert an input token sequence  into a feature vector. TWIN and SIAMESE only need $O(N*K)$ to convert artifacts into feature vectors and then $O(N^2)$ time to concatenate the feature vectors together. The time complexity of TWIN and SIAMESE is one order of magnitude lower than the SINGLE model thus more scalable to projects with massive artifacts. 
We argue that SIAMESE is the most appropriate model for addressing NLA-PLA traceability taking both accuracy and efficiency into consideration, because it can achieve an accuracy close to the SINGLE architecture for the traceability challenge while maintaining the low time complexity of the TWIN architecture. However, in cases where accuracy is the primary concern, e.g. traceability for safety critical projects, users should adopt SINGLE model supported by high-performance hardware.

\noindent{$\bullet$ \bf RQ1.3:} How do T-BERT models compare to other approaches ?

For the CodeSearchNet challenge, we compared the performance of T-BERT models to Joint Vector Embedding (JVE) and MS-CodeBERT. JVE's architecture is similar to TWIN, and leverages two encoders to create feature vectors for a classification network. Previous studies have reported 60\% as the highest  MRR achieved by JVE on the same dataset, which is lower than T-BERT models. MS-CodeBERT, provided by Microsoft, used the same architecture as SINGLE in our experiment. However,  MS-CodeBERT was trained with a batch size of 256 on a cluster with 16 Tesla GPU, and no special techniques were applied during training. Our machine only allows one small batch due to memory limitations, but SINGLE* model's MRR results were only 0.9\% lower than MS-CodeBERT, indicating that our training techniques  partially alleviate limitations introduced by less powerful hardware. 

\begin{figure}[tbh!]
    \centering
    \includegraphics[width=\linewidth]{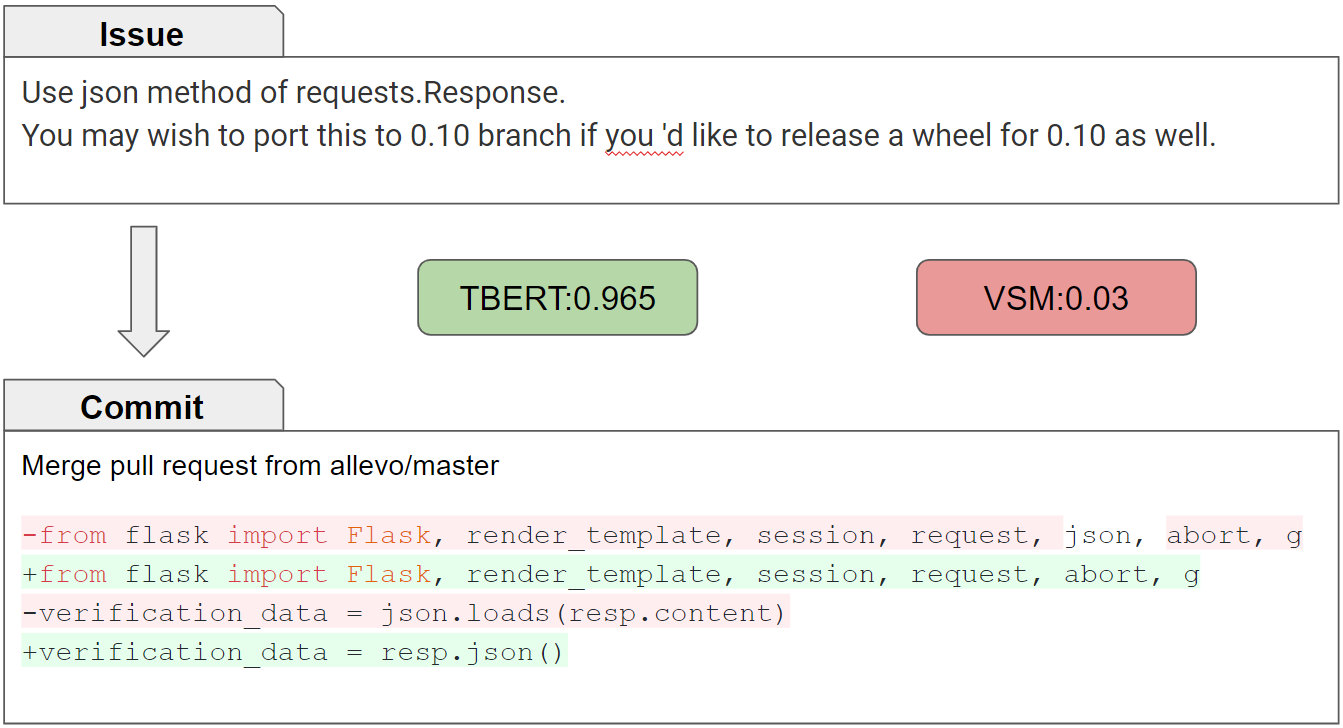}
    \caption{
    In this example, a link is tagged by developers,  retrieved by the T-BERT model (with a high score o 0.965 due to semantic similarity and context) and missed by VSM, because the key terms  'request` and 'json` are common terms.}
    \label{tb_fig:trace_example}
\end{figure}

\def\TraceNN{TNN }
\TraceNN \cite{guo2017semantically} is an RNN based trace model proposed by Guo et al. and designed for generating NLA-NLA links. We reconstructed the model according to the authors' specifications and applied it to our NLA-PLA problem for comparison purposes. \TraceNN utilizes Word2Vec embedding to transform tokens into vectors. It uses two alternate RNN networks, LSTM or Bi-directional GRU (BiGRU) to generate semantic representations of NLA and PLA, and feeds these semantic hidden states to the integration layer to generate a new hidden state representing the correlations between the NLA and PLA from which links are generated. Our embedding layer was constructed by unsupervised training of a skip-gram Word2Vec model using artifacts from the three OSS reported in Table \ref{tb_tab:oss_data}. We evaluated both LSTM and BiGRU for the RNN layer in this study.
\TraceNN results are shown at the bottom of Table.~\ref{tb_tab:trace_acc}, and show that it underperformed all BERT models and VSM on all three OSS projects. We provide an illustrated example in Fig.~\ref{tb_fig:trace_example}, showing T-BERT and VSM results for a commit-issue pair, tagged by the committer as related.

Guo et al., reported improvements over the VSM model for their NLA-NLA dataset, we were unable to replicate these for our NLA-PLA problem.
An inspection of the \TraceNN learning curve indicated that \TraceNN effectively reduced the loss and improved the link prediction accuracy for all three training datasets in training dataset, but converged early in the validation datasets and then decreased in accuracy - indicating an overfitting problem. There are several possible explanations for these results. First, the dataset used by Guo et al., contained 1,387 positive links versus our 530-739 links, which could be insufficient for RNN training. Second, programming languages have an open vocabulary in which new terms can be created as variable and function names, and \TraceNN may therefore need a larger training set to generate NLA-PLA links versus NLA-NLA ones. Our hypotheses are supported by the observation that \TraceNN does not overfit when applied to the CodeSearchNet where larger numbers of training examples are provided.
T-BERT models leverage transferred knowledge from pretrained language models and adjacent problems to reduce the requirements of the training dataset size, and are therefore able to handle tracing challenges which can not easily be addressed by classical Deep Learning trace models. This characteristic makes T-BERT more practical for industrial applications.


\begin{figure}[t!]
    \centering
    \includegraphics[width=\linewidth]{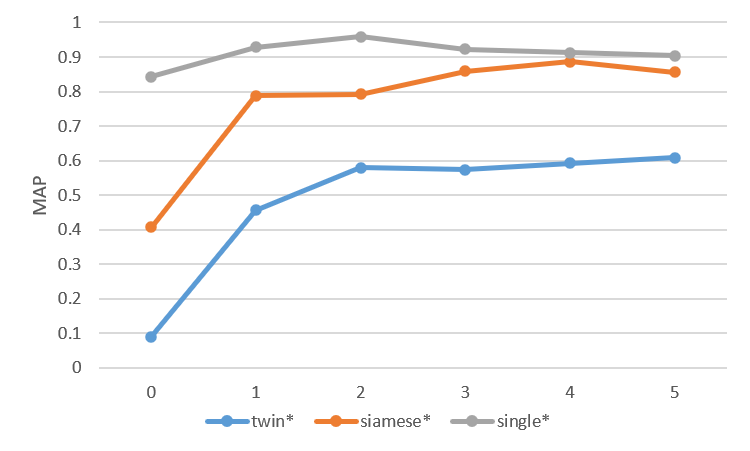}
    \caption{Learning curve of T-BERT* models on Pgcli dataset for NLA-PLA trace challenge. This figure shows the MAP scores (Y-Axix) over 5k Adam optimization steps (X-Axis)}
    \vspace{-6pt}
    \label{tb_fig:trace_curve}
\end{figure}

\subsection{RQ3: To what extent can T-BERT leverage transfer knowledge from code search to software traceability}
Table ~\ref{tb_tab:trace_acc} the T-BERT model trained with and without transferred knowledge from the post-pretrained model. The results show that \posttraining T-BERT models on the code search task can significantly improve their performance on the traceability problem. Taking SIAMESE on Pgcli for example, the F2 score increased from 0.654 to 0.748, while the MAP score increased from 0.728 to 0.779. Similar results are observed for the other datasets with different T-BERT models. This suggests that the knowledge learned from text to structure code (function definition) can be effectively transferred to cases where 1) code formats are more fuzzy and 2) training data has limited labels. 
\Posttraining improvements  were observed to different extents across the three architecture types. As shown in Fig.~\ref{tb_fig:trace_curve}, the blue line (SINGLE) converged at a very early stage, showing that SINGLE needed only relatively few epochs on the smaller task-specific dataset to localize its transferred knowledge. SIAMESE converged slower, while TWIN converged slowest of all, indicating that each architecture has a different capacity for transferring knowledge.

\section{Related Work}
\label{tb_sec:related_work}
Our study constructs T-BERT models using three different architectures, all of which have previously been used to address related problems in other domains. Lu et al. leveraged a TWIN-like model, named TwinBERT, as a search engine to deliver ads alongside  organic search results \cite{lu2020twinbert}. They used reinforcement training techniques and found that TwinBERT could return results with high accuracy and low latency. 
Reimers et al. proposed a SIAMESE architecture to address problems such as Semantic Textual Similarity \cite{reimers2019sentence}.  They trained their model to determine whether two sentences were related through contradiction or entailment, and utilized SNLI\cite{snli:emnlp2015} and the Multi-Genre NLI \cite{N18-1101} dataset for training and evaluation. Their results showed that SIAMESE BERT could achieve a high Spearman Rank Correlation score around 0.76. They also found that use of averaging pooling was more effective than max pooling and first token (the [CLS] token) pooling; and that concatenating source and target hidden states as $(u,v,\|u-v\|)$ achieved best results. We adopted these findings when we built T-BERT models for this study.  However, no study has yet been conducted comparing the TWIN, SIAMESE and SINGLE architectures.

To address the NLA-PLA challenge, we adopted the code search problem as our intermediate solution. Several studies have addressed the code search problem using a recurrent neural network (RNN). We have already made comparisons to the work  by Feng et al. \cite{feng2020codebert} and Huian et al. \cite{husain_codesearchnet_2019} in Table.~\ref{tb_tab:csn_acc}; however, in another study, Gu et al. \cite{DBLP:conf/icse/GuZ018} converted method specifications into API call sequences and then processed the sequence with RNN. They reported achieving  0.6 MRR in a test set with 100 queries. However, we can not directly adapt this method to the traceability challenge, because, unlike API calls, the statements in code change sets are not structured. 
A related domain for addressing NLA-PLA is source code embedding. By converting both source code and documents into distributed representations, the relevance between these two type of artifacts can be effectively calculated through distance metrics such as Cosine and Euclidean Distance. Code2Vec \cite{alon2019code2vec} belong to this type of approach. T-BERT models can adapt to this type of training by integrating Cosine Embedding Loss in the classification header. We leave this exploration for future work. 

\section{Threats to Validity}
There are several threats to validity in this study. 
First, Our current experiments have only been conducted on Python projects, and results could differ when  applied to other programming languages. Also, due to time constraints we fine-tuned and evaluated the T-BERT model performance on only three OSS projects, which may not be enough to draw generalized conclusions.
Second, we construct our experiment datasets from OSS projects by mining the issues and commits whose IDs are explicitly  marked as related by project maintainers. Although, this is a conventional way of leveraging OSS projects for traceability, true links may be missed. For example, a bug report may have hidden dependencies on several other issues such as a feature request or other bug report even though a commit addressing the parent bug report is not marked as `related'. We alleviate the impact of this phenomena by adopting the data processing suggested by Liu et.al. \cite{liu2020traceability}.
Another important threat is that while the SINGLE architecture, trained for code search problem, does not outperform CodeBERT, further improvements could be achieved using hyper parameter optimization. Our experiments were limited by hardware availability for conducting excessive hyper parameter tuning. However, the performance comparison across T-BERT models should still be valid because all experiments were conducted with the same parameters. Finally, due to processing time constraints, we evaluated the SINGLE model on 100 queries whilst using the  entire testing set for the other models (in Table.~\ref{tb_tab:csn_acc}). Although not reported we also evaluated TWIN and SIAMESE on 100 queries and observed that they achieved almost identical results to those obtained from the whole test set indicating that 100 queries was a reasonable sample size for the SINGLE model.

\section{Conclusion and Future Work}
\label{tb_sec:conclusion}
This study has explored several different BERT architectures for generating trace links between natural language artifacts and programming language artifacts. Our experimental results showed that the SINGLE architecture achieved the best accuracy but at long execution times, whilst the SIAMESE architecture achieved similar accuracy with faster execution times. 
Second, we showed that ONS training (based on negative sampling) improved both performance and model convergence speed without incurring significant performance overheads when compared to DRNS. Third, we found that T-BERT was able to effectively transfer knowledge learned from the code search problem to NLA-PLA traceability, meaning that \posttrained T-BERT models can be effectively applied to software engineering projects with limited training examples, alleviating the data sparsity problem for deep neural trace models. Regarding the training time, we showed that the same \posttrained T-BERT can be applied for OSS projects in three different domains. By avoiding the need for intermediate training on each individual project, our approach was able to efficiently adapt to new domains.
In conclusion, our results  show that T-BERT generates trace links at far higher degrees of accuracy than existing  information retrieval and RNN techniques -- bringing us closer to achieving the vision of practical and trustworthy traceability. 

To support replication and reproducibility, we have provided links throughout this paper to the datasets that we used and we provide a complete implementation of T-BERT and execution instructions on github\footnote{T-BERT source code https://github.com/jinfenglin/TraceBERT}.

In future work we will evaluate our approach across more diverse project domains and programming languages, and will explore its application to more diverse types of software artifacts such as requirements, design, and test cases.


\balance
\section*{Acknowledgment}
This work has been partially funded under US National Science Foundation Grant SHF:1901059.

\bibliographystyle{./bibliography/IEEEtran}
\bibliography{./bibliography/ICSE2021, ./bibliography/trace}

\end{document}